\documentclass[pra,twocolumn,showpacs,
floatfix, reprint]{revtex4-2}
\usepackage{graphicx}

\usepackage{amsmath}

\usepackage{hyperref}
\hypersetup{
  colorlinks   = true, 
  urlcolor     = blue, 
  linkcolor    = blue, 
  citecolor   = blue 
}

\begin{document}


\title{Narrow-line cooling of $^{87}$Rb using 5S$_{1/2} \rightarrow$ 6P$_{3/2}$ open transition at 420 nm}


\author{Rajnandan Choudhury Das}
 \author{Dangka Shylla}%
 \author{Arkapravo Bera}
\author{Kanhaiya Pandey}%
\email{kanhaiyapandey@iitg.ac.in}
\affiliation{%
 Department of Physics, Indian Institute of Technology Guwahati, Guwahati, Assam 781039, India
}%

\date{\today}

\begin{abstract}
Magneto-optical trap (MOT) at narrow (weak) transition offers lower temperature and hence is the key for production of high phase density atomic cloud and subsequently quantum degeneracy with high number of atoms for many elements. In this paper, we describe loading of $^{87}$Rb atoms in the MOT using a narrow open transition at 420 nm from the routinely implemented MOT using broad cyclic transition at 780~nm (IR). The total linewidth of the blue transition, 5S$_{1/2} \rightarrow $ 6P$_{3/2}$ is 1.4 MHz, which is around 4 times narrower than the standard 5S$_{1/2} \rightarrow$ 5P$_{3/2}$ cyclic transition. Using this narrow transition, we have trapped around $10^{8}$ atoms in the MOT with a typical temperature of around $54~\mu$K. We have also studied the behavior of the blue MOT with various parameters such as hold time, detuning and power of trapping and repumper beams.
\end{abstract}

\maketitle

\section{\label{sec:lev1}Introduction}
In Doppler cooling, lower linewidth of transition is desirable in order to achieve lower temperature, as the Doppler temperature (T$_D$) is proportional to the linewidth of the transition ($\Gamma$) given by the equation T$_D=\hbar\Gamma/2k_B$, where $\hbar$ is the reduced Planck's constant and $k_B$ is the Boltzmann constant. However capture velocity in the MOT being proportional to the linewidth, makes the broader linewidth transition favorable for capturing higher number of atoms. Hence, in order to achieve efficient cooling with large number of atoms, two stage MOT has been used for many elements such as Ca \cite{Hollberg2001, Hollberg2003}, Sr \cite{Makoto1999, Wilk2014EPJD},  Yb \cite{Yabuzaki1999, Natarajan2010}, Dy \cite{Benjamin2011, Pfau2014},  Er \cite{Jabez2008, Ferlaino2012, Boong2020}, Cd \cite{Katori2019}, Eu \cite{Mikio2021} etc.

Alkali atoms have transition linewidth of around 2$\pi\times$6~MHz for D$_2$ transition which corresponds to Doppler temperature of around 150 $\mu$K. Temperature can further be reduced to few tens of $\mu$K for Na, Rb and Cs using the polarization-gradient cooling mechanism. 
However this sub-Doppler cooling mechanism demands switching off the MOT magnetic field for few ms which leads to the expansion of the cloud. This makes the cloud density smaller and inefficient for loading into the optical dipole trap (ODT) for evaporative cooling.
Hence the MOT at narrow transition is quite useful for efficient transfer to the ODT. For narrow transition, the levels with different principal quantum numbers can be utilized, as the electric dipole matrix element between them is small due to less overlap of the wave-functions, which causes the transition to be weak (narrow).

For the two alkali atoms Li and K, polarization-gradient cooling does not work because of the closely spaced hyperfine levels of the excited states and to achieve sub-Doppler temperature, gray molasses on D$_1$ line \cite{Christophe2013,Fernandes2012,Unnikrishnan2013,Salomon2013,Pan2016} and D$_2$ line \cite{Bruce2017} have been used. But again for gray molasses, the MOT magnetic field has to be switched off for few ms leading to the expansion of the cloud. The narrow line cooling has been implemented in Li \cite{Hulet2011, Dieckmann2014} and K \cite{Thywissen2011}. Further its advantage over gray molasses for loading the atoms into ODT has been demonstrated in Li \cite{Dieckmann2018}.     
Rb also possesses similar level structure as that of Li and K. However, demonstration of reduction in temperature of $^{87}$Rb MOT using narrow line transition is important because of its complex decay channels, specially when the repumper laser is at broad transition. 

In this work, we characterize the loading of the blue MOT of $^{87}$Rb atoms at 420 nm (5S$_{1/2}$, F$=2 \rightarrow$ 6P$_{3/2}$, F$=3$) from the MOT at 780 nm (5S$_{1/2}$, F$=2 \rightarrow$ 5P$_{3/2}$, F$=3$). For both the MOTs, we use only one repumper laser at 780 nm (5S$_{1/2}$, F$=1 \rightarrow$ 5P$_{3/2}$, F$=2$) to pump the atoms back to the cooling cycle from the other ground state hyperfine level (5S$_{1/2}$, F$=1$).

\section{\label{sec:lev2}Experimental Set-up}
\subsection{Laser system and spectroscopy}\label{spectro}

\begin{figure}[t]
  \includegraphics[width=1\linewidth]{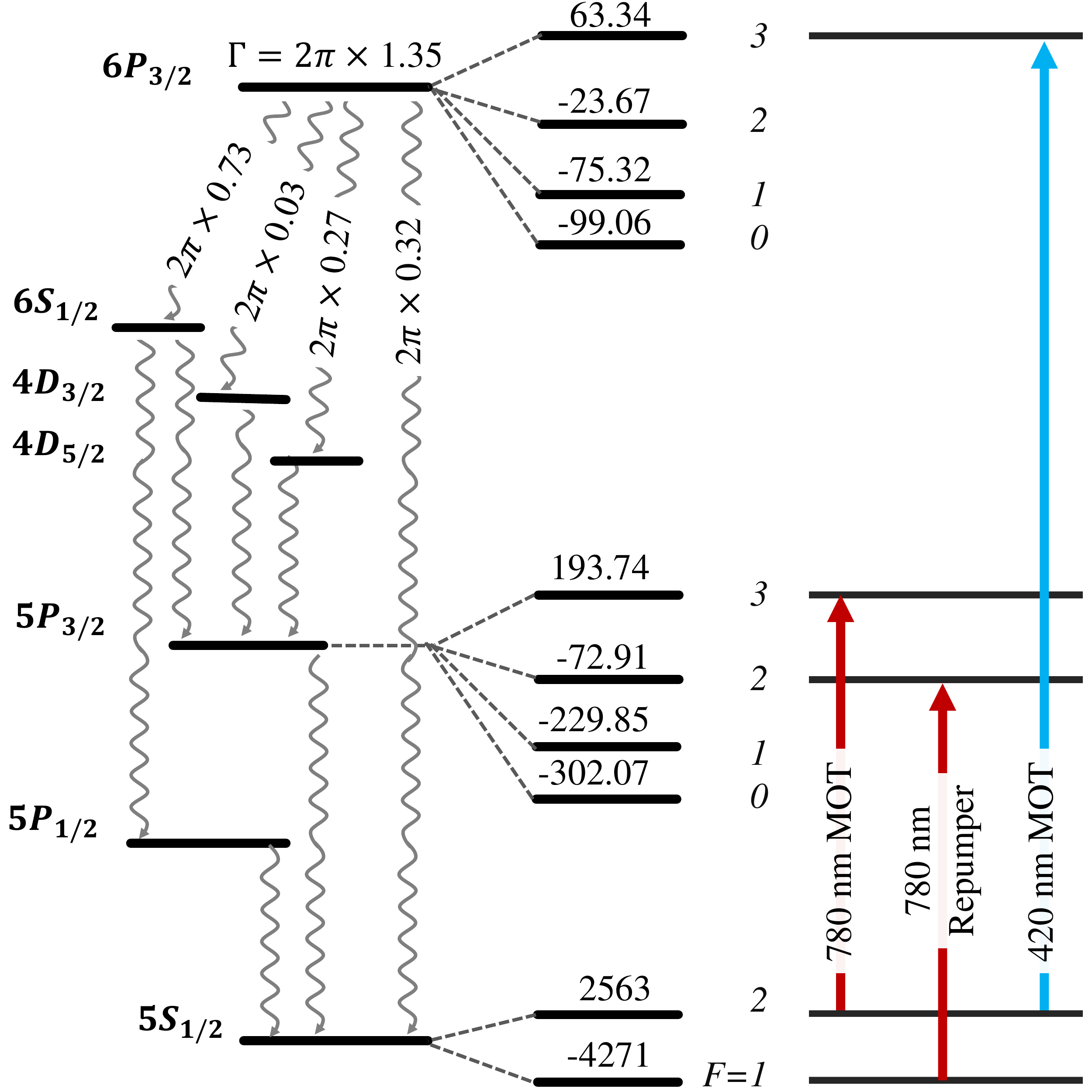}
  \caption{\label{EnergyLevels}(Color online) The relevant energy levels of $^{87}$Rb with the hyperfine splittings and various decay paths of the 6P$_{3/2} $ state. Decay rates, linewidth of the excited state and the hyperfine splittings are shown in MHz unit.}
  \end{figure}

The relevant energy level diagram of $^{87}$Rb along with the hyperfine splittings and its complex decay paths of the 6P$_{3/2}$ state is shown in Fig. \ref{EnergyLevels}. Hyperfine splittings of 5S$_{1/2}$, 5P$_{3/2}$ and 6P$_{3/2}$ are adopted from \cite{Jozsef2020} and decay rates of various decay channels of the 6P$_{3/2}$ state are calculated using the dipole matrix elements given in \cite{Clark2004}. Calculated linewidth of 6P$_{3/2}$ is $2\pi\times1.35$~MHz and its corresponding lifetime is 118~ns \cite{Clark2004}.

The 780 nm (IR) lasers are generated from two separate home-assembled external cavity diode lasers (ECDL). Both the 780 nm trapping and repumper lasers use laser diodes (make: Thorlabs, model: L785H1 and L785P090 respectively). The output beam diameter is $2~\text{mm}\times 3~\text{mm}$ for both the IR lasers and the output power available are $73 \text{ mW}$ and $13 \text{ mW}$ respectively. The 420 nm (blue) laser is generated from a commercially available ECDL (make: TOPTICA, mode: DL PRO HP) with a typical linewidth $<200~\text{kHz}$, output power 70 mW, and beam diameter $3~\text{mm}\times 4~\text{mm}$.

\begin{figure}[t]
  \includegraphics[width=1\linewidth]{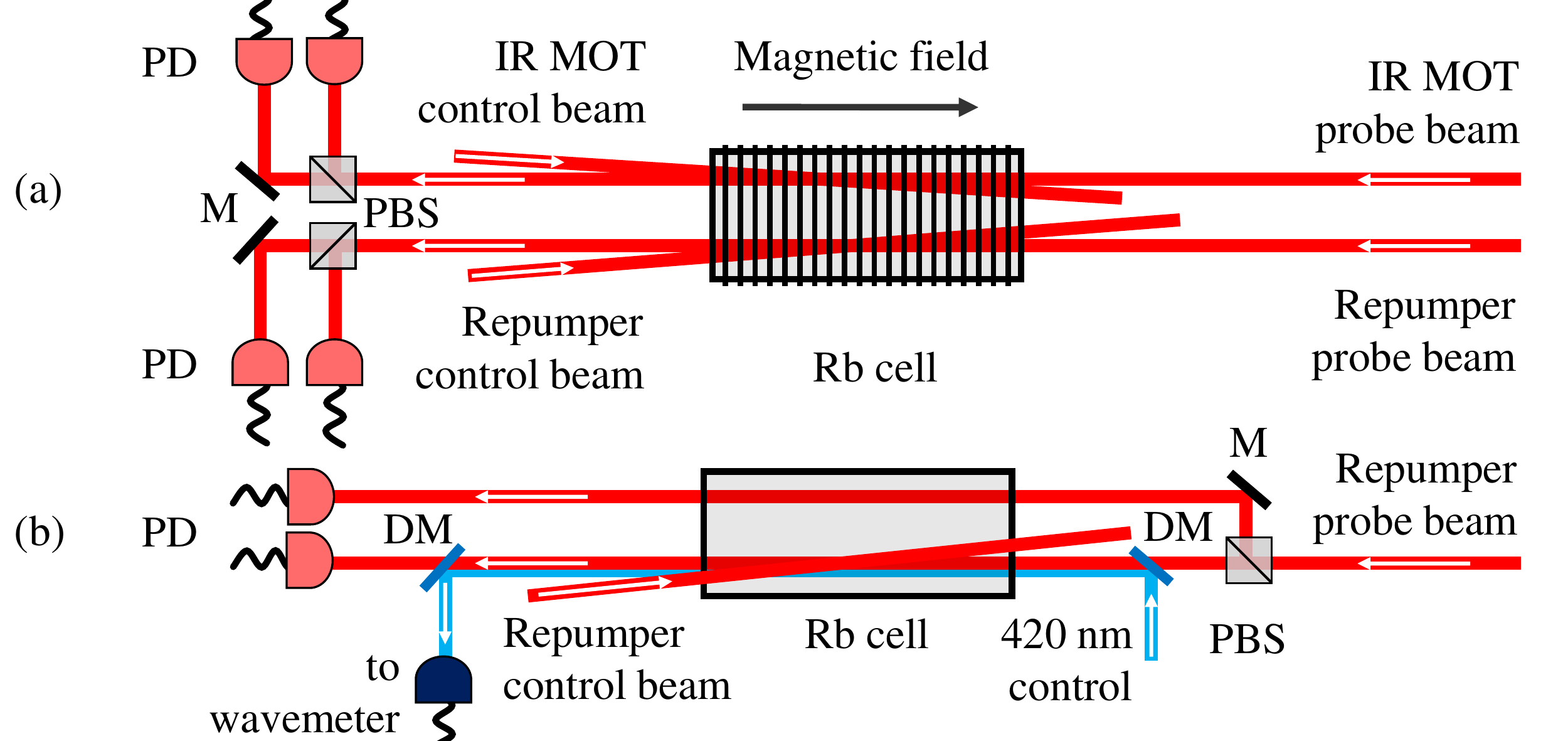}
  \caption{\label{fig:exp_setup}(Color online) (a) Polarization spectroscopy scheme for 780 nm MOT and repumper lasers. (b) Double resonance spectroscopy scheme for 420 nm laser. Figure abbreviations: DM: dichroic mirror, M: mirror, PBS: polarizing beam splitter, PD: photo-detector. 780 nm and 420 nm beams are shown in red and blue color respectively.}
\end{figure}

The 780 nm MOT laser is divided into three parts. First part is utilized for frequency stabilization using polarization spectroscopy technique \cite{Hughes_JPB_2008} as shown in Fig. \ref{fig:exp_setup} (a). The Rb vapor cell is heated up to 50 \textdegree C inside an oven using two peltier coolers to improve the signal to noise ratio of the absorption spectrum. Peltier coolers are placed beneath the cell closer to its edge to avoid coating on the windows.  Power of the control and probe beams are around $500~\mu$W and $300~\mu$W respectively. The laser is locked to the 5S$_{1/2}$, F $= 2 \rightarrow$ 5P$_{3/2}$, F $= (2,3)$ crossover peak of the $^{87}$Rb absorption spectrum. Its second part is sent through an acousto-optic modulator (AOM) at frequency $+2\times66.5$~MHz in double pass configuration and is used for imaging the MOT cloud after cleaning the spatial modes to a Gaussian. This AOM is used mainly for three purposes: firstly, for fast switching of the imaging beam, secondly for scanning the imaging beam frequency to find the resonance of the MOT beam, and thirdly to change the detuning of the imaging beam. Double pass configuration is chosen as it prevents misalignment when detuning of the imaging beam is changed. The third part is used as trapping beam for the IR MOT after frequency shifting by $+123.5$~MHz in single pass configuration through AOM. This AOM is used for fast switching and power variation of the 780 nm trapping beam.

The 780 nm repumper laser is also divided into three parts. First part is sent to the same vapor cell where the spectroscopy of the 780 nm MOT laser is done. Separation between the IR MOT probe beam and repumper probe beam is around 7-8 mm. The repumper laser is frequency stabilized at 5S$_{1/2}$, F $= 1 \rightarrow$ 5P$_{3/2}$, F $= 2$ transition again using polarization spectroscopy \cite{Hughes_JPB_2008} as shown in Fig. \ref{fig:exp_setup} (a). The power of the control and probe beams are $2$~mW and $25~\mu$W respectively. The second part is sent to the MOT chamber after mixing it with the MOT beams. The third part is sent as probe beam to another Rb vapor cell for double resonance spectroscopy of the 420 nm laser with similar scheme as \cite{Pandey_JPB_2020,Pandey_PRA_2021}.

\begin{figure}[t]
  \includegraphics[width=1\linewidth]{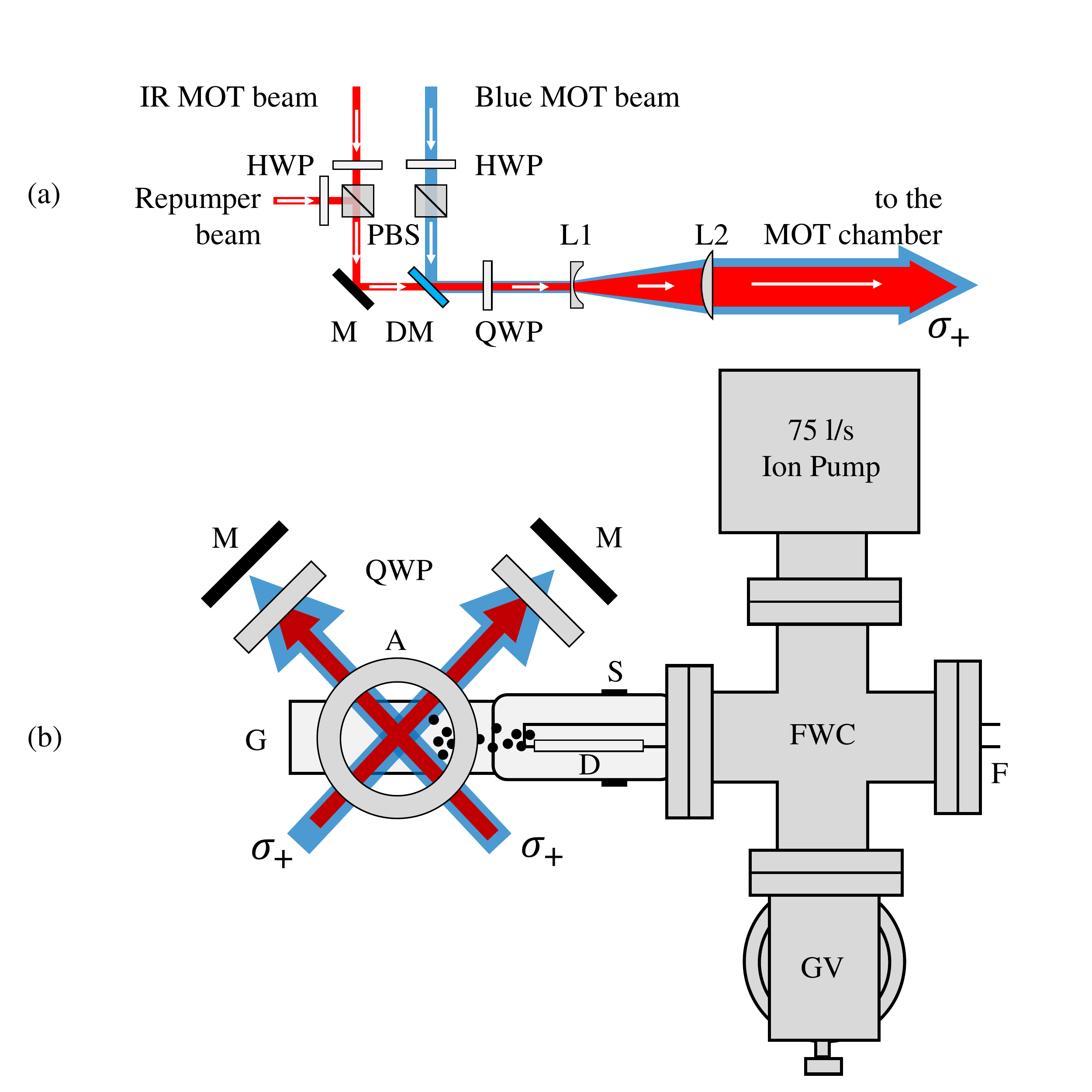}
  \caption{\label{fig:exp_setup_2}(Color online) (a) Mixing scheme of the IR MOT, repumper and blue MOT beams. (b) Top view of the MOT set-up. Figure abbreviations: A: anti-Helmholtz coil, D: Rb dispenser, DM: dichroic mirror, F: electric feedthrough, FWC: 4-way cross, G: glass chamber, GV: all metal gate valve, HWP: $\lambda/2$ wave-plate, L$_{1}$: plano-concave lens, L$_{2}$: plano-convex lens, M: mirror, PBS: polarizing beam splitter, S: Quartz to metal seal, QWP: dual $\lambda/4$ wave-plate, $\sigma_{+}$: co-propagating MOT beams in $\sigma_{+}$ configuration. 780 nm and 420 nm beams are shown in red and blue color respectively.}
\end{figure}

The 420 nm laser is divided into two parts. First part is sent to the second vapor cell through an AOM at frequency $-2\times88.0$~MHz in double pass configuration and is used as a control (with power of around $800~\mu$W) in the co-propagation configuration with 780 nm repumper laser which is stabilized at 5S$_{1/2}$, F$=1\rightarrow$ 5P$_{3/2}$, F$=2$ transition. The IR repumper laser acts as a probe and is monitored at photo-detector as shown in Fig. \ref{fig:exp_setup} (b). The power of this IR laser is around $100~\mu$W. In order to remove intensity noise of the laser and  improve the signal to noise ratio of the double resonance spectrum, we use another reference IR beam which passes through the same cell and falls on another photo-detector. Both the photo-detectors are in A$-$B configuration. The error signal to lock the blue laser is generated by modulating the AOM at 50 kHz and it is locked to the 5S$_{1/2}$, F $= 1 \rightarrow$ 6P$_{3/2}$, F $= 2$ transition. The same AOM is also used for changing the detuning of the 420 nm trapping beam. Single mode operation and the wavelength of the 420 nm laser is monitored using a wavelength meter (make: Highfinesse GmbH, model: WS7-60). Second part of the blue beam is frequency up-shifted by 93.0 MHz using an AOM in the single pass configuration and is sent to the mixing scheme, where it is overlapped and co-propagated with the 780~nm MOT and repumper beams as shown in Fig. \ref{fig:exp_setup_2} (a) before sending it to the MOT chamber. This AOM is also used for fast switching and power variation of the 420 nm trapping beam.

\subsection{MOT set-up}
The 780 nm trapping, repumper and 420 nm trapping beams are overlapped and co-propagated in the mixing scheme as shown in Fig. \ref{fig:exp_setup_2} (a). Three beams are derived in such a way that polarization of the trapping beams are identical in each of the three arms. Then the three beams are made circularly polarized using dual $\lambda /4$ wave-plates and are expanded individually by 10 times using Galilean telescopes consisting of a pair of plano-concave and plano-convex lens of focal length, f = $- 25$ mm and $250$ mm respectively. The expanded beams are then sent to the MOT chamber and retro-reflected back with the combination of a mirror and a  dual $\lambda /4$ wave-plate.

The schematics of the MOT set-up is shown in Fig. \ref{fig:exp_setup_2} (b). The MOT chamber used in this work is a rectangular glass chamber with dimension $7.5~\text{cm}\times 2.5~\text{cm}\times 2.5~\text{cm}$. It is connected to an ion pump (pumping speed 75~l/s) through a 4-way cross. The pressure inside the vacuum chamber is around 10$^{-10}$ mbar. Atomic rubidium vapor is dispensed into the chamber by passing 2.2 A current to a dispenser (AlfaSource Rubidium, model: AS-Rb-0090-2C-RbBi40) through the electric feedthrough at other end of the 4-way cross. This dispenser current is kept fixed throughout the work except for the lifetime measurement where dispenser current is varied from 1.8 A to 2.4 A. The distance between the opening mouth of dispenser to the MOT center is around 6~cm.

One pair of coils each having inner diameter 7~cm, outer diameter 13~cm, thickness 3 cm and 500 number of turns in an anti-Helmholtz configuration is used to create magnetic-field gradient of around 18 Gauss/cm at the center of the MOT chamber by passing current, I = 1 A. Three pairs of shim coils surrounding the MOT chamber are also used (not shown in the schematics) to null the stray magnetic-field. Fast switching of the magnetic field is done using an insulated-gate bipolar transistor (IGBT) based high current switch. The switching off time of the coil is around $20~\mu$s.

\begin{figure}[t]
  \includegraphics[width=1.0\linewidth]{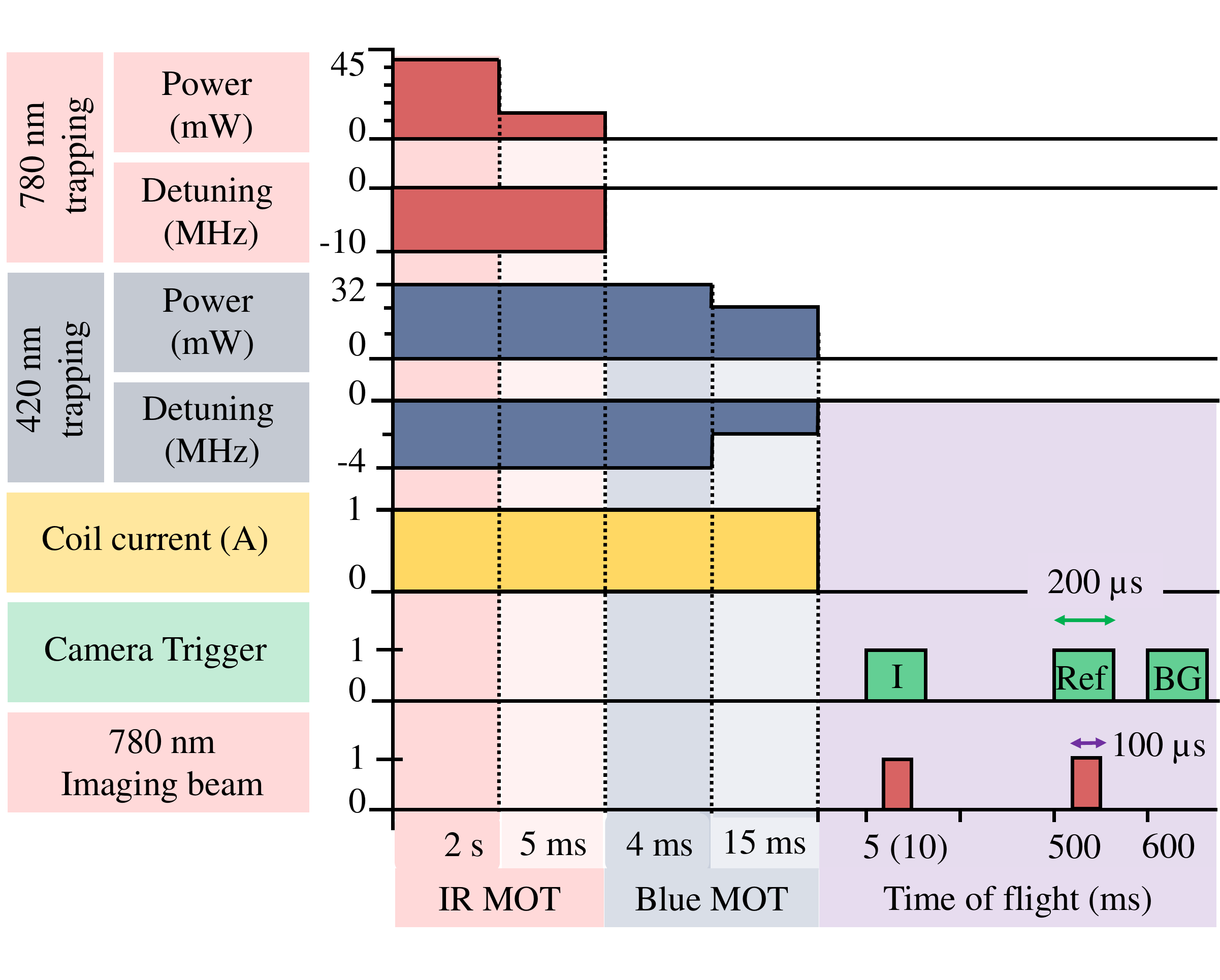}
  \caption{\label{fig:TimeSeq}(Color online) Time sequence along with the corresponding experimental parameters for loading the IR and blue MOT and characterization of the MOT cloud. The coil current of 1 A corresponds to the magnetic field gradient of around 18~Gauss/cm. The ON (OFF) state of the camera trigger and 780 nm imaging beam is referred as 1 (0). Three images captured at each cycle (i) image with MOT cloud, (ii) reference beam, and (iii) background are shown as I, Ref, and BG respectively. The axes are not to scale.}
\end{figure}

Time sequence along with the corresponding experimental parameters for loading the IR and blue MOT as well as the characterization of the MOT cloud are shown in Fig. \ref{fig:TimeSeq}. Various parameters of the MOT such as number of atoms, number density, temperature, and lifetime are determined by performing the absorption imaging of the MOT at 5S$_{1/2}$, F$=2 \rightarrow$ 5P$_{3/2}$, F$=3$ close transition using the imaging beam derived from the 780 nm MOT laser as explained in the section \ref{spectro}.
Power of the imaging beam is around 3 $\mu$W and its diameter is 20 mm (peak intensity 0.002~mW/cm$^2$). The imaging beam is switched on only for 100 $\mu$s during imaging.

Images are captured on a CMOS camera (make: Thorlabs, model: CS135MUN) using 1-f imaging system of de-magnification 0.3342. Exposure time of the camera is kept at 200 $\mu$s.
Three images (i) image with MOT cloud, (ii) Reference beam, and (iii) Background are captured using the time sequence shown in Fig. \ref{fig:TimeSeq} to determine the size of the cloud and the number of atoms. Same sequence is executed twice but with two different time of flights (TOF) at 5~ms and 10~ms, to measure the temperature of the MOT.

\section{Results}
\subsection{IR MOT}
First, the atoms are loaded in an IR MOT using 780 nm trapping beam at 45 mW power and -10 MHz detuning for 2 s. In order to lower the temperature of the IR MOT, we lower the power of the 780 nm trapping beam to 10 mW for 5 ms using the time sequence as shown in Fig. \ref{fig:TimeSeq}. The parameters are set by conducting an experiment using a little bit different set-up of the 780 nm MOT laser in master-slave configuration.
In this set up, the master laser is locked to the atomic transition and a part of the beam of this laser is up-shifted using an AOM in double pass configuration and used for seeding the slave laser. The slave laser remain seeded for the AOM frequency variation of around 13 MHz which allows us to vary the MOT laser detuning by 26 MHz.
We observe that at the given parameters, $\sim10^{8}$ number of atoms are trapped in the IR MOT at typical temperature of around 400 $\mu$K. The loading time of the IR MOT is around 1.5~s.

\subsection{Blue MOT}
\begin{figure}[t]
    	\includegraphics[width=1\linewidth]{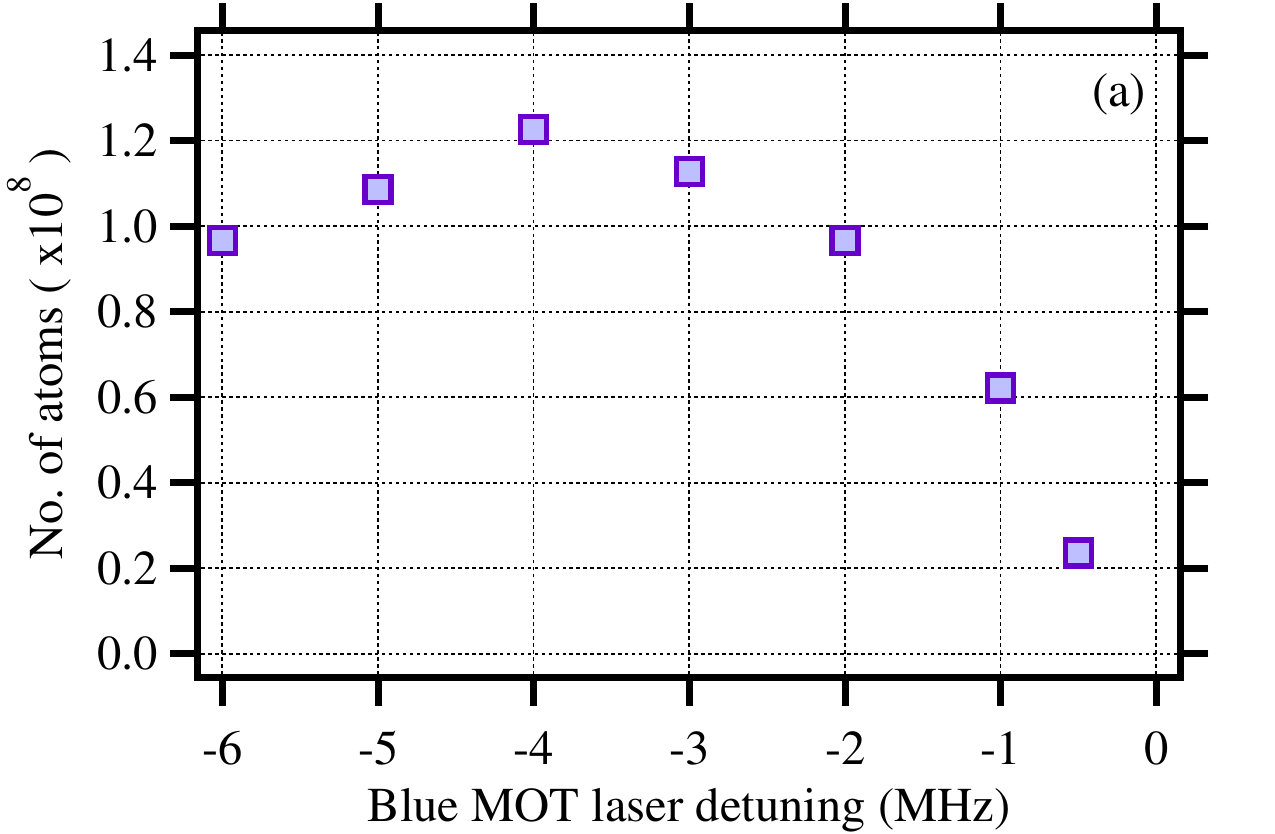}
	\hfill
 	\includegraphics[width=1\linewidth]{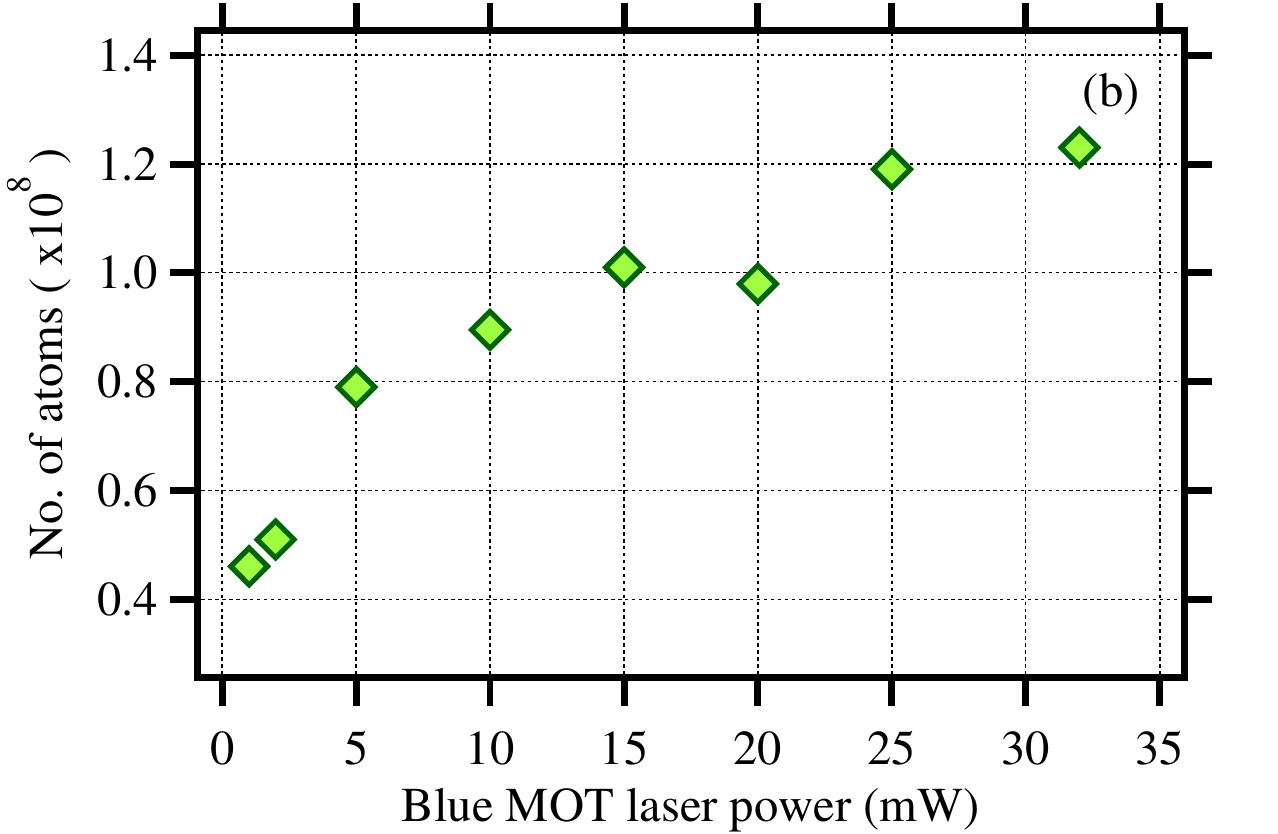}
	\hfill
	\includegraphics[width=1\linewidth]{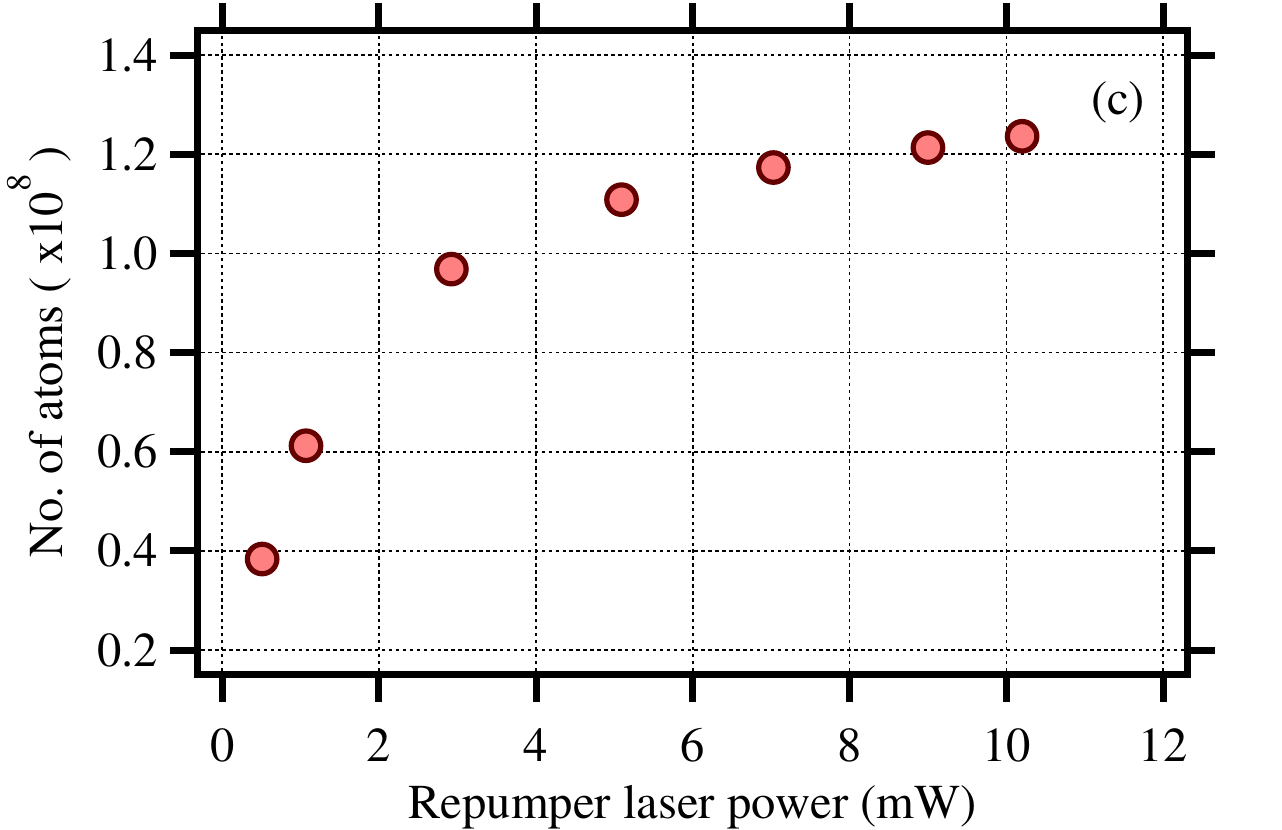}
    	\caption{\label{fig:No_power}(Color online) Number of atoms is shown as a function of (a) 420 nm MOT laser detuning in loading phase, (b) total 420 nm MOT laser power and (c) 780 nm repumper power. In (a) and (b) repumper laser is kept at its maximum available power, 10.2 mW. In (b) and (c) detuning of the 420 nm trapping beam in loading phase is kept at -4 MHz. In (a) and (c) power of the 420 nm MOT beam is kept at its maximum available power, 32 mW.}
 \end{figure}

Atoms are then loaded into the blue MOT by turning off the IR MOT beam, but keeping the blue MOT beam on for 4 ms at maximum available power of the 420 nm trapping (32~mW) and repumper (10.2~mW) beams. Number of atoms transferred to the blue MOT is monitored for various detuning of the blue MOT laser in the loading phase at this fixed power and is shown in Fig. \ref{fig:No_power} (a). From the plot, it is found that the number of atoms transferred to the blue MOT increases as the detuning of the blue MOT laser is changed from $-6$~MHz to $-4$~MHz and then decreases as the detuning is changed towards the resonance. At $-4$~MHz detuning of the 420 nm trapping beam, maximum number of atoms ($1.2\times 10^{8}$) are transferred to the blue MOT. This detuning is hence used  throughout the work for loading the blue MOT.

Next, to see the effect of the power of the blue MOT laser, we vary its power and monitor the number of atoms in the blue MOT. Detuning of the blue MOT is kept at $-4$~MHz and the power of the repumper beam is kept at maximum. From the Fig. \ref{fig:No_power} (b), it is clear that the number of atoms first increases with increase in power of the blue MOT laser and then saturates at high power. It seems that 32~mW power of the blue MOT laser is good enough for the efficient transfer and is used throughout the work for loading the blue MOT.

To check the required power of the repumper, we again monitor the number of atoms in the blue MOT vs the repumper power as shown in Fig. \ref{fig:No_power} (c). The power of the blue MOT laser is kept at maximum and its detuning is kept at $-4$~MHz. We observe similar pattern as that with the blue MOT laser power. Number of atoms first increases with increase in repumper power and then saturates to a maximum at high power of the repumper beam. We observe that 10.2 mW of repumper power is good enough to capture $1.2 \times10^{8}$ atoms in the blue MOT.

\begin{figure}[t]
  \includegraphics[width=1\linewidth]{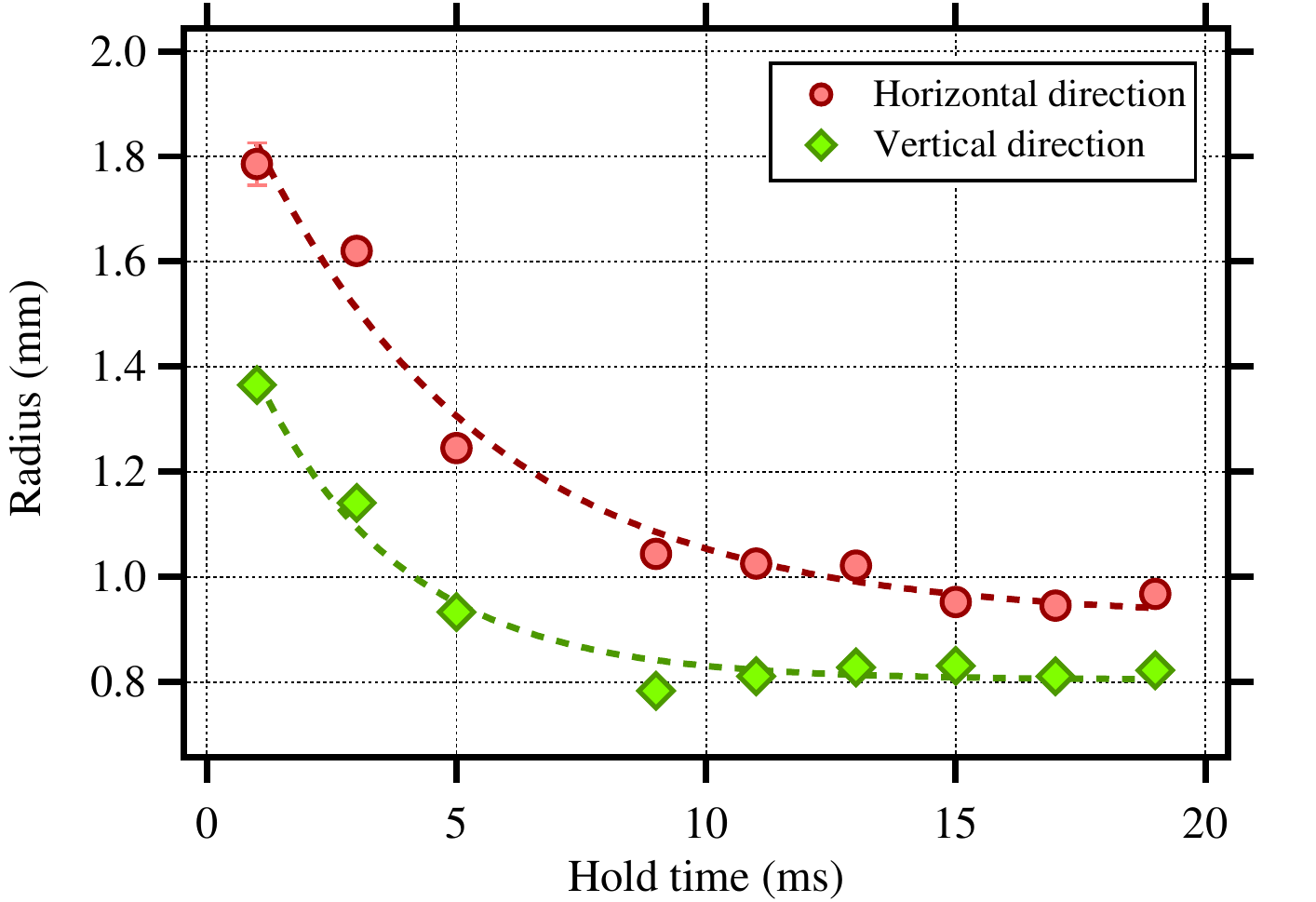}
  \caption{\label{fig:radius} (Color online) Radius of the cloud along horizontal (red circle) and vertical (green diamond) directions versus hold time in the blue MOT are shown for short hold time. Exponential fits are shown as dashed lines, with 1/e time of 4.7 ms and 2.9 ms for horizontal and vertical direction respectively. Images were captured at 0.5 ms time of flight after release from the trap. }
\end{figure}

After transferring the atoms from the IR MOT to the blue MOT, we lower the power of the blue MOT laser and vary the hold time and the detuning of the blue MOT laser. For fixed power and detuning, we observe that the radius of the blue MOT decreases exponentially with the hold time, signifying the cooling of the MOT cloud. Fig. \ref{fig:radius} shows its typical variation with the hold time at 10 mW power and $-2$~MHz detuning. The radii data are fitted to exponential decay function and it is found that blue MOT size equilibrate with 1/e relaxation time of 4.7~ms and 2.9~ms for horizontal and vertical direction respectively. It is also clear from the graph that the MOT cloud does not change its size typically after 15~ms hold time and hence the power of the blue laser is lowered for 15~ms for further studying the effect of blue laser detuning on number of atoms and temperature.

\begin{figure}[t]
   	\includegraphics[width=1.0\linewidth]{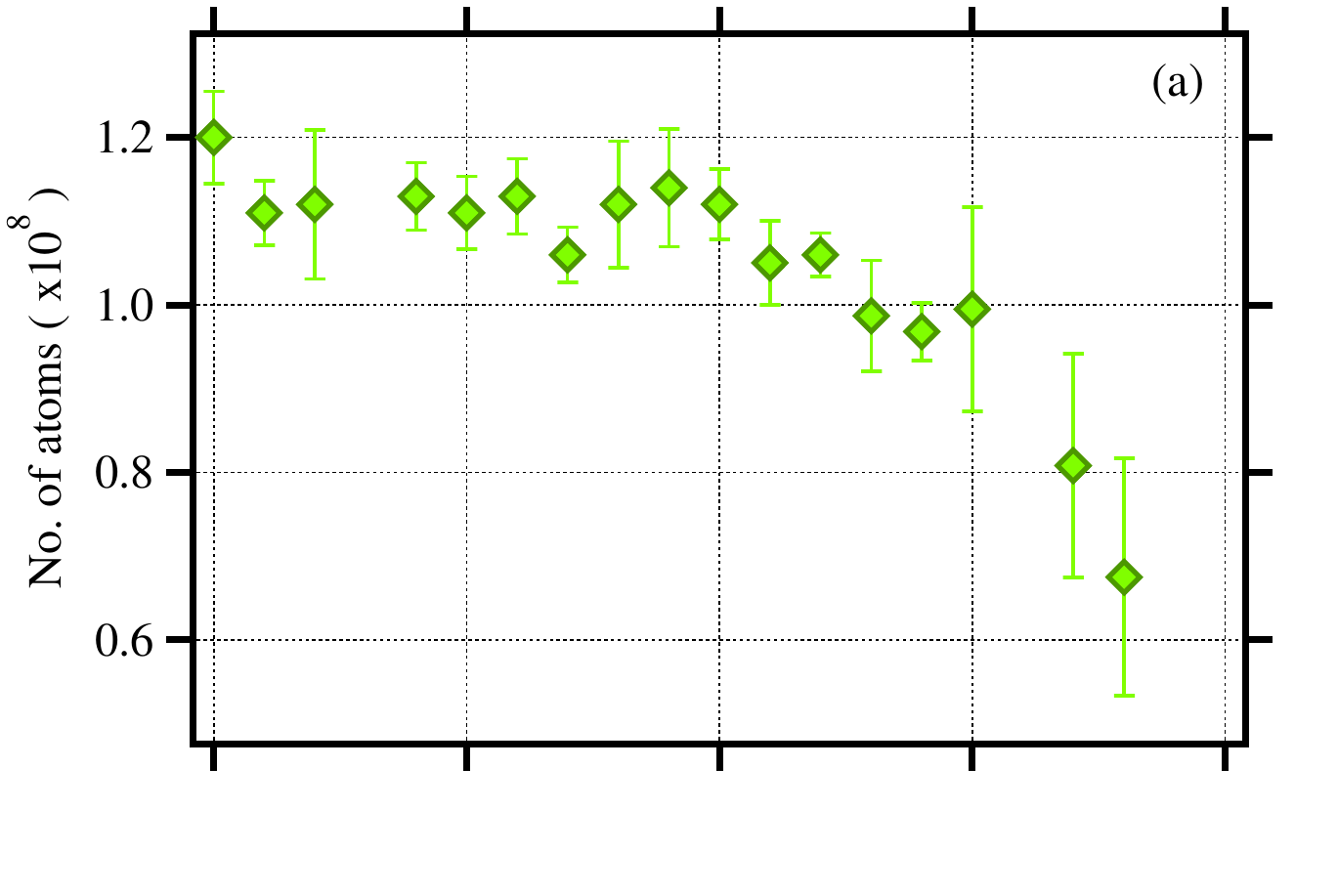}
	\hfill
	\includegraphics[width=1.0\linewidth]{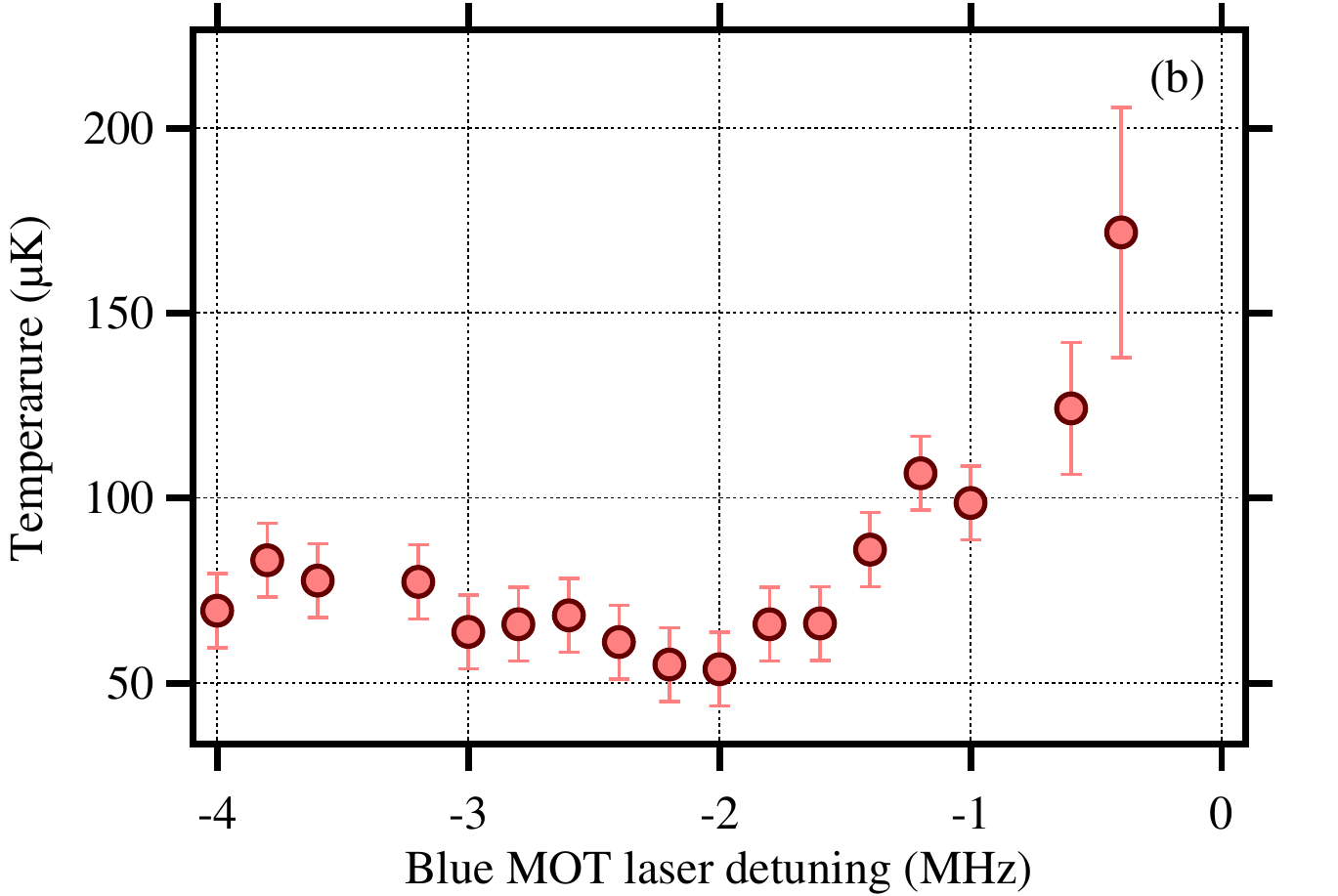}
	\caption{\label{fig:Detuning}(Color online) (a) No. of atoms, and (b) Temperature of the Blue MOT are shown as a function of the blue MOT laser detuning. Number of atoms (green diamond) are probed at 5~ms time of flight and temperatures of cloud (red circle) are measured from time of flight measurement at 5~ms and 10~ms.}
 \end{figure}

Next we fix the hold time as 15~ms, vary the detuning of the blue MOT laser at 10 mW and study the behavior of the blue MOT by monitoring the the number of atoms and temperature of the blue MOT. Number of atoms are measured by probing the MOT cloud using the time sequence shown in Fig. \ref{fig:TimeSeq} at TOF 5~ms and its variation with detuning are shown in Fig. \ref{fig:Detuning} (a).We observe that the number of atoms remains almost constant ($\sim1.1 \times 10^{8}$) in the range of $-4$~MHz to $-2$~MHz of the detuning of the blue MOT laser and then decreases as the detuning is changed towards zero.

Temperature of the cloud is determined  by measuring the cloud size at two different TOF using the same time sequence executed twice but with TOF 5~ms and 10~ms. Its variation with detuning of the blue MOT laser is shown in Fig. \ref{fig:Detuning} (b). We observe that at $-2$ MHz, the temperature of the blue MOT is minimum and it is around $54(10)~\mu$K. Temperature increases slowly to around $80~\mu$K as the deturning is further increased towards $-4$ MHz from the point of minimum, and it increases rapidly to around $170~\mu$K as the detuning is changed towards resonance.

Number of atoms in the blue MOT at the minimum temperature of $54~\mu$K is $\sim1.1 \times 10^{8}$ and the corresponding number density is $\sim1.0 \times 10^{10}$~cm$^{-3}$.

\begin{figure}[t]
  \includegraphics[width=1\linewidth]{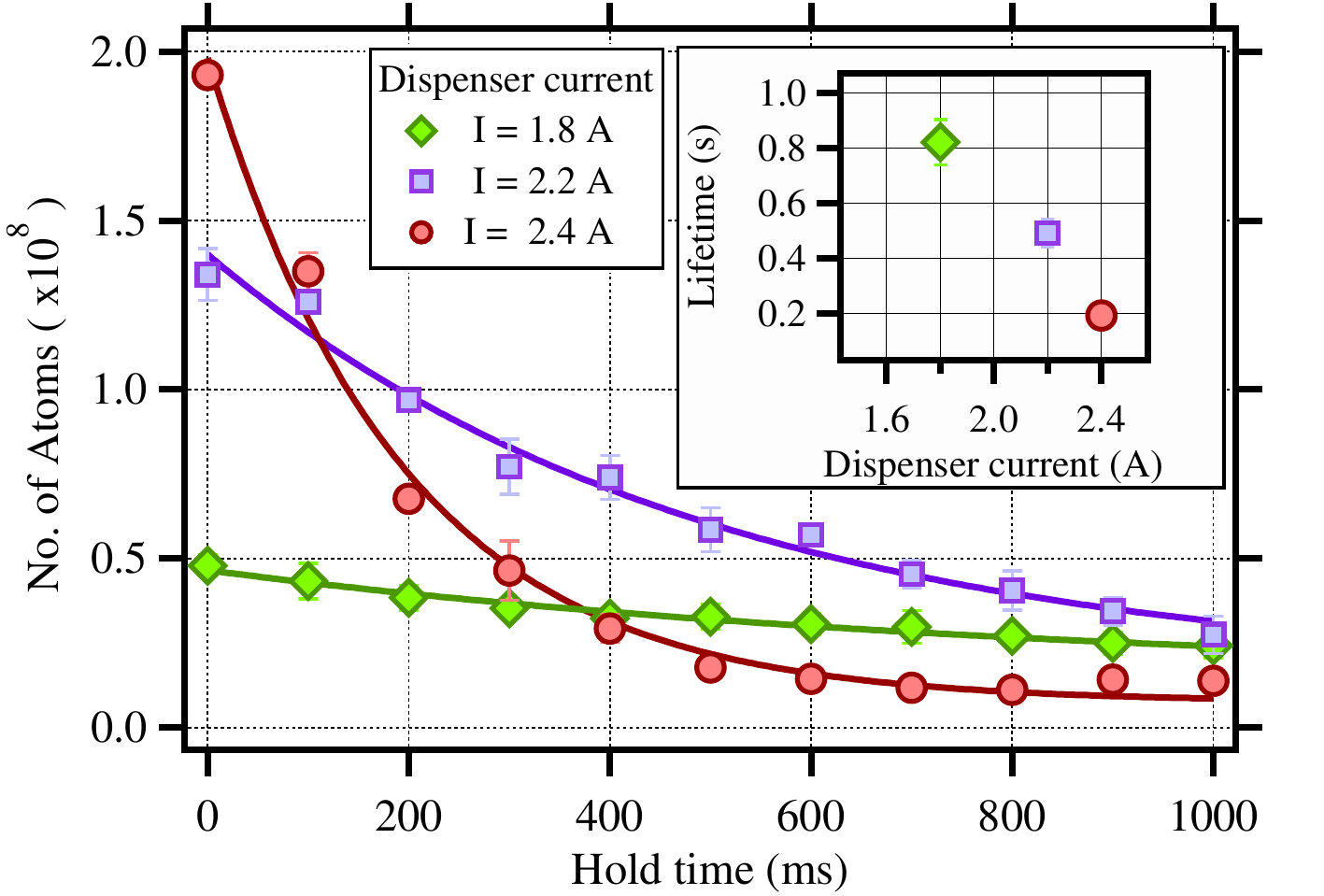}
  \caption{\label{fig:Lifetime}(Color online) Lifetime of the blue MOT. No. of atoms is shown versus the hold time in the blue MOT at three dispenser current: I = 1. 8 A (green diamond), I = 2.2 A (violet square), and I = 2.4 A (red circle). The data are fit to exponential decays (solid lines) and the corresponding lifetimes are shown in the top inset. No. of atoms are probed at 0.5 ms TOF. }
\end{figure}

To determine the lifetime of the blue MOT, we monitor the number of atoms in the blue MOT at different hold time. We observe that the number of atoms decreases exponentially with hold time as shown in Fig. \ref{fig:Lifetime}. The lifetime of the blue MOT is determined by fitting the data with exponentially decay function. It is found that the lifetime of the blue MOT is around $0.5$~s at dispenser current of $2.2$~A.

We also study the effect of dispenser current on the number of atoms in the blue MOT and its lifetime.
A trade off between the maximum number of trapped atoms and the lifetime of the blue MOT is observed. It is found that with increase in dispenser current, lifetime of the MOT decreases as shown in Fig. \ref{fig:Lifetime} (inset), although the maximum number of atoms in the blue MOT increases. Maximum of $5 \times 10^{7}$ atoms can be trapped in the blue MOT with maximum lifetime of 0.8 s at 1.8 A dispenser current. 4 fold increase in number of trapped atoms to $2 \times 10^{8}$ are achieved by increasing the dispenser current to 2.4 A at the expense of decrease in lifetime by around 4 times to 0.2 s. This is because of the increase in collision with the background atoms due to increase in flux of atoms when dispenser current is increased.

\section{\label{sec:lev4}Conclusion}
In summary, we have demonstrated the narrow-line cooling of $^{87}$Rb atoms in the MOT using the open transition, 5S$_{1/2} \rightarrow$ 6P$_{3/2}$ at 420 nm to trap around $1.1\times10^{8}$ number of atoms at typical temperature of $54(10)~\mu$K. Even though the branching ratio is small ($\sim1/4$), we have achieved efficient cooling in the blue MOT, which may further be improved by using a 420 nm repumper laser in place of the broad 780 nm repumper laser.

\begin{acknowledgments}
RCD would like to acknowledge Ministry of Education, Government of India for the Prime Minister's Research Fellowship (PMRF). K.P. would like to acknowledge the funding from DST through Grant No. DST/ICPS/QuST/Theme-3/2019.
\end{acknowledgments}


\bibliography{Blue_MOT}

\end{document}